\documentclass[useAMS,usenatbib]{mn2e} 
\usepackage{epsfig}
\usepackage{color}

\newcommand{\Msolar}{\mbox{\,$\rm M_{\odot}$}} 
\newcommand{\Rsolar}{\mbox{\,$\rm R_{\odot}$}} 
\def\simge{\mathrel{\raise1.16pt\hbox{$>$}\kern-7.0pt \lower3.06pt\hbox{{$\scriptstyle\sim$}}}} 
\def\simle{\mathrel{\raise1.16pt\hbox{$<$}\kern-7.0pt \lower3.06pt\hbox{{$\scriptstyle \sim$}}}} 

\newcommand{\iso}[2]{\mbox{$^{#1}{\rm #2}$}} 

\begin{document}

\title[R CrB stars]{Can R\,CrB stars form from the merger of two helium white dwarfs?} \author[]{Xianfei
Zhang$^{1}$\thanks{E-mail: xiz@arm.ac.uk} and C. Simon Jeffery$^{1,2}$\thanks{E-mail:
csj@arm.ac.uk}\\ $^1$Armagh Observatory, College Hill, Armagh BT61 9DG, UK\\ $^2$School
of Physics, Trinity College Dublin, Dublin 2, Ireland }

\date{Accepted . Received ; in original form }

\pagerange{\pageref{firstpage}--\pageref{lastpage}} \pubyear{2011}

\maketitle

\label{firstpage}

\begin{abstract} Due to orbital decay by gravitational-wave radiation, some close-binary
helium white dwarfs are expected to merge within a Hubble time. The immediate merger
products are believed to be helium-rich sdO stars, essentially helium main-sequence
stars. We present new evolution calculations for these post-merger stars beyond the core
helium-burning phase. The most massive He-sdO's develop a strong helium-burning shell and
evolve to become helium-rich giants. We include nucleosynthesis calculations following
the merger of $0.4 \rm M_{\odot}$ helium white-dwarf pairs with metallicities $Z =
0.0001, 0.004, 0.008$ and 0.02. The surface chemistries of the resulting giants are in
partial agreement with the observed abundances of R Coronae Borealis and extreme helium
stars. Such stars might represent a third, albeit rare, evolution channel for the latter, in addition
to the CO+He white dwarf merger and the very-late thermal pulse channels proposed
previously. We confirm a recent suggestion that lithium seen in R\,CrB stars
could form naturally during the hot phase of a merger in the presence of \iso{3}{He} from the donor white dwarf.
\end{abstract}


\section{Introduction}
The existence of close binary white dwarfs has been predicted theoretically \citep{Han98,
Nelemans00, Nelemans01} and demonstrated observationally \citep{Rebassa11, Brown11,
Kilic10, Kilic11}. It has been demonstrated that the orbits of close binaries decay
through the emission of gravitational wave radiation \citep{Evans87, Cropper98}. Models
for the merger of nearly-equal mass white dwarfs following spiral-in demonstrate that
these occur on a dynamical timescale (minutes) \citep{Benz90, Guerrero04, Pakmor11}.
The less massive WD is disrupted and part of the debris forms a prompt hot corona, the remainder forms a disk which subsequently
accretes onto the surviving white dwarf \citep{Yoon07,Loren09}. Models for the evolution
of the product of a double helium-white-dwarf merger demonstrate the off-centre ignition
of helium burning, followed by expansion and evolution onto the helium main sequence
\citep{Saio00}. Models which include the composite nature of the debris (corona + disk)
successfully account for the distribution in effective temperature, luminosity and
surface composition of compact helium-rich subdwarf O stars \citep{Zhang12}. In
particular, the most-massive mergers become the hottest sdO stars with carbon-rich
surfaces, whilst the least-massive mergers are cooler with predominantly nitrogen-rich
surfaces.

Following inward migration of the helium-burning shell, the merged star is essentially a
helium main-sequence star having a mass between 0.4 and 1.0 M$_{\odot}$, although its
surface layers will carry a record of the merger event. The evolution of low-mass helium
stars has been studied for over 40 years \citep{Paczynski71, Dinger72, Trimble73,
Weiss87}. The lowest-mass stars will evolve directly to become CO/He white dwarfs
following core helium exhaustion. For $M \ge1.0 {\rm M}_{\odot}$, shell helium burning
will ignite around the carbon/oxygen core, and the star will expand to become a giant.
Early authors proposed this as a possible origin for the hydrogen-deficient R Coronae
Borealis (R CrB) variables \citep{Paczynski71, Weiss87}.

R CrB stars and the hotter extreme helium stars are low-mass supergiants of spectral
types F, A and B. Their surfaces are extremely deficient in hydrogen, and enriched in
carbon, oxygen, neon and other nuclear waste. Two principle evolution channels have been
established. A small fraction may be produced following a late thermal pulse in a
post-asympotic giant-branch star on the white dwarf cooling sequence \citep{Iben84a,
Clayton11}. The majority are more likely to have been produced following the merger of a
carbon-oxygen white dwarf with a helium white dwarf \citep{Webbink84, Saio02, Jeffery11,
Longland11}.

In this letter we demonstrate that high-mass double helium-white-dwarf mergers could
contribute a third evolution channel and account for some, at least, of the lowest
luminosity R CrB and extreme helium stars.


\section{Models}
\subsection{Evolution}
Numerical simulations of stellar evolution are
carried out using the stellar-evolution code "Modules for Experiments in Stellar
Astrophysics" (MESA) \citep{paxton11}. Helium-white-dwarf models were generated by
evolving a 2$\Msolar$ main sequence star and removing the envelope when this reached the
required mass (see \citet{Zhang12} for details). The post-merger evolution was modelled
by considering fast accretion at $10^4 \rm \Msolar yr^{-1}$ (representing the formation
of a hot corona), followed by slow accretion at $10^{-5} \rm \Msolar yr^{-1}$
(representing accretion from a Keplerian disk). \citet{Zhang12} considered three sets of
assumptions concerning this accretion: i) all fast, ii) all slow, and iii) composite. In
the composite model, most of the mass of the donor white dwarf is ingested into the hot
corona, with $0.1 \rm \Msolar$ remaining in the Keplerian disk for slow accretion.

The evolution from off-center helium ignition to the end of core helium burning has been
discussed and compared favourably with observations of helium-rich subdwarf B and O stars
\citep{Zhang12} . Following core helium exhaustion, the post-merger models with $M=0.5,
0.6$ and $0.7\Msolar$ developed a brief phase of thick helium shell burning, before
helium burning ends and the stars contract to become CO/He white dwarfs. The models with
$M=0.8\Msolar$ ($M_{\rm core}= 0.64 \Msolar$) developed a thin helium-burning shell and
evolved to high luminosity and low effective temperature. This is a regime occupied by
another class of hydrogen-deficient star, the R\,CrB stars.

We have made additional calculations for these 0.4+0.4 \Msolar mergers
with metalicities $Z = 0.0001, 0.004, 0.008$ and 0.02 to investigate whether
the He+He WD merger channel could produce some of the observed EHe and R\,CrB stars.

\subsection{Nucleosynthesis}
In the SPH simulations of white dwarf mergers, matter in the hot coronae may
briefly reach temperatures of $6\rm\times10^8 K$ or more \citep{Yoon07,Loren09}, so that
some nucleosynthesis of $\alpha$-rich material will occur. Similarly, our one-dimensional
quasi-equilibrium calculations indicate coronal temperatures up to $4\rm\times10^8 K$ for
a 0.4+0.4 \Msolar merger \citep{Zhang12} . At these temperatures, the 3$\alpha$ and other
alpha capture reactions are ignited at the surface of the accretor almost immediately.
Thus, carbon is produced by helium burning and nitrogen is destroyed by $\alpha$ capture.
As \citet{Warner67} and \citet{Clayton07} indicated, the destruction of \iso{14}{N} by
the $\iso{14}{N}(\alpha,\gamma)\iso{18}{O}$ reaction becomes more efficient than it is at
low temperature. As the temperature continues to increase, \iso{18}{O} produces
\iso{22}{Ne} through $\alpha$ capturing. The question is whether the composition of these
surface layers also matches that of the R\,CrB stars.


\begin{figure}
\centering \includegraphics [angle=0,scale=0.48]{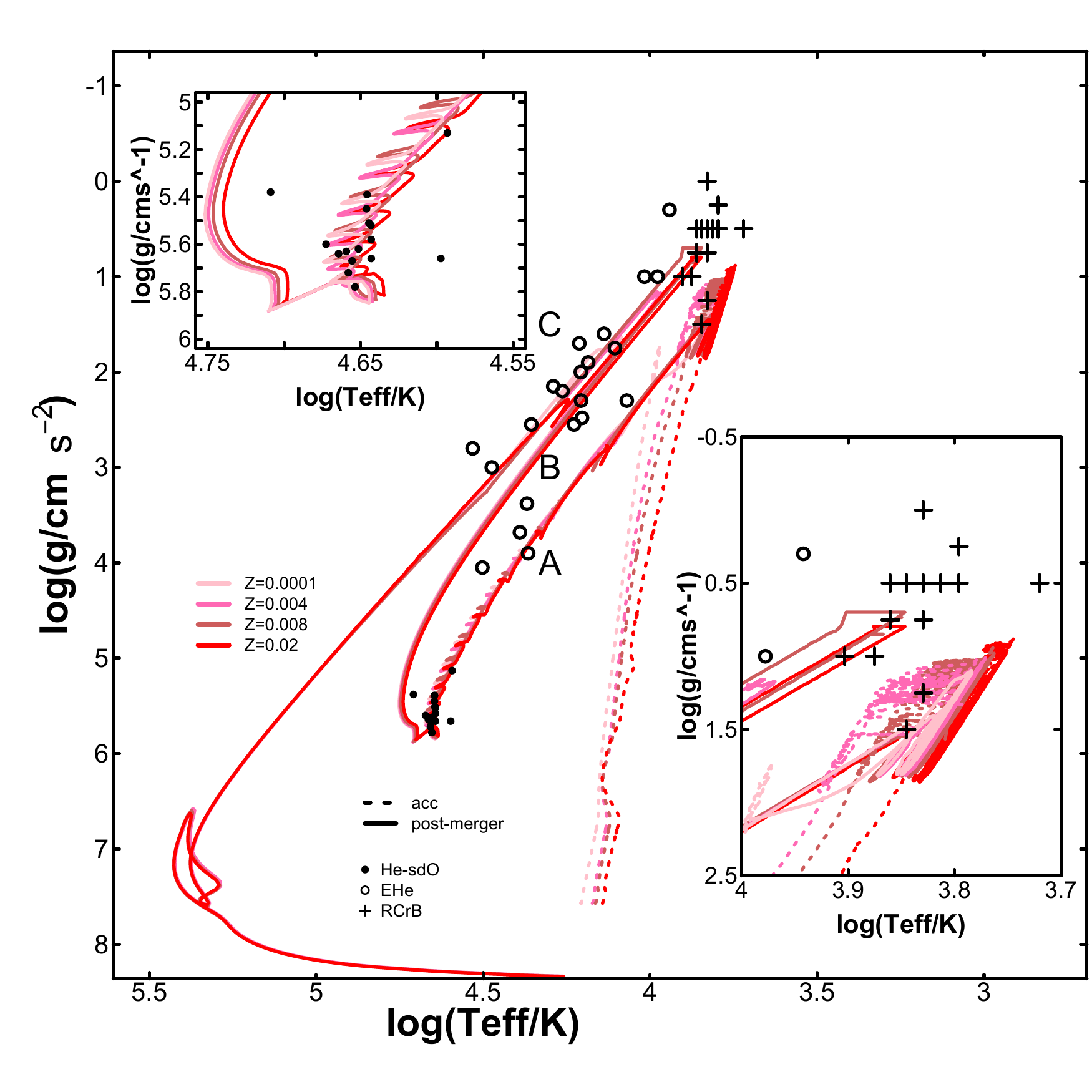}
\caption{Evolutionary tracks at different metallicities on a surface gravity -- effective
temperature diagram for a 0.8 $\Msolar$ post-merger model for a He+He white-dwarf binary.
The dashed line shows the evolution track during accretion, solid lines show the
post-merger evolution. The colour from grey to dark (a colour version is
available in the online journal, from pink to red) shows the different metallicities,
i.e. $Z=0.0001, 0.004, 0.008, 0.02$. The cross symbols shows the R\,CrB stars
from \citet{Jeffery11}, the circle symbols shows the EHe stars from
\citet{Jeffery11}, the dots show the He-sdO stars from \citet{Hirsch09}. The inset top
left shows an enlargement around the phase of core helium burning.
The inset lower right shows an enlargement of the R\,CrB star region.
 Three stages are marked:
A for  inward shell-burning;
B for outward
shell-burning; C for giant branch to white dwarf.  }
\label{gt}
\end{figure}

\section{Results} \subsection{Evolution tracks}

Taking the $Z= 0.02$ model as an example, during the fast accretion stage
a hot corona with  a radius of $0.06 \Rsolar$ is formed within 16 minutes, whilst the luminosity ($\log L/{\rm L_{\odot}}$)
increases from -2 to -1.  As the fast accretion
starts, the coronal temperature reaches $10^8$K very quickly. At the end of the fast phase,
the helium-burning shell reaches a peak temperature of about $4\times 10^8$K. During this
process, an almost completely convective envelope is produced due to the corona being heated by
the energy produced by helium buring.

After the fast merger phase, the remaining mass of the secondary forms a Keplerian
disc surrounding
the accretor and transfers mass to the central object at a rate comparable to the Eddington
accretion rate. During this slow-accretion phase, the helium-burning shell still heats the
corona, forcing it to continue expanding. Thus, the star expands to $\approx 32 \Rsolar$
within $4\times 10^3$ yrs and then contracts away from the giant branch (Fig.~\ref{gt}).

During subsequent evolution, there follow about 20 helium flashes in $6\times10^5$ yrs, each
subsequent flash decreasing in intensity, until the helium-burning shell (flame)
reaches the centre of the star. This corresponds to the end of the loops on the $g-T{\rm eff}$
diagram. After this, a standard core-helium burning
phase is established, followed by a normal helium-shell burning phase, leading to higher
luminosity, and finally cooling to the white dwarf sequence. Just before reaching
the white dwarf sequence, there is a small loop caused by a weak final helium shell flash (Fig.~\ref{gt}).

In order to compare with observations, data for 13 R\,CrB stars and 21 EHe stars are taken from
\citet{Jeffery11} and for 15 He-sdO stars from \citet{Hirsch09}. As Fig.~\ref{gt} shows, most
of the EHe and R\,CrB star lie around the low-gravity evolution corresponding to outward
shell-helium burning after  core-helium burning.  Four or five of the EHe stars lie closer
to the high-gravity  inward shell-burning phase. After
this phase they evolve to become He-sdO stars as \citet{Saio00} indicated.
Two R\,CrB stars (RT Nor and RS Tel) have high enough gravities to be immediate post-merger
objects.  They may still have a disc surrounding them or be surrounded by dust.
Thus, there is a channel from the the He+He white dwarf merger which can produce R\,CrB stars,
then possibly EHe stars, and then He-sdO stars. Subsequently, massive He-sdO stars might
evolve to become EHe and R\,CrB stars once again, before finally cooling to become white dwarfs.
We note that this channel will not generate metal-poor R\,CrB stars, due to lower opacity in the
stellar envelope, but can generate low-metallicity EHe stars (e.g. BD+10 2179: Kupfer et al. 2012).


\subsection{Surface composition}

At the beginning of the fast-accretion phase, the corona is fully convective, and
helium burning makes the corona rich in $\iso{12}{C}$, $\iso{18}{O}$ and $\iso{22}{Ne}$.
At the end of fast accretion and beginning of slow accretion, flash-driven convection
mixes \iso{12}{C} throughout the 0.3$\Msolar$ envelope. Although this material is
subsequently buried by C-poor material from the disk, the deep opacity-driven convection which follows
each shell-pulse dredges carbon-rich material to the new surface, so that the final
product, after slow-accretion terminates, is rich in \iso{12}{C} and \iso{22}{Ne}.

\begin{figure} \centering \includegraphics [angle=0,scale=0.45]{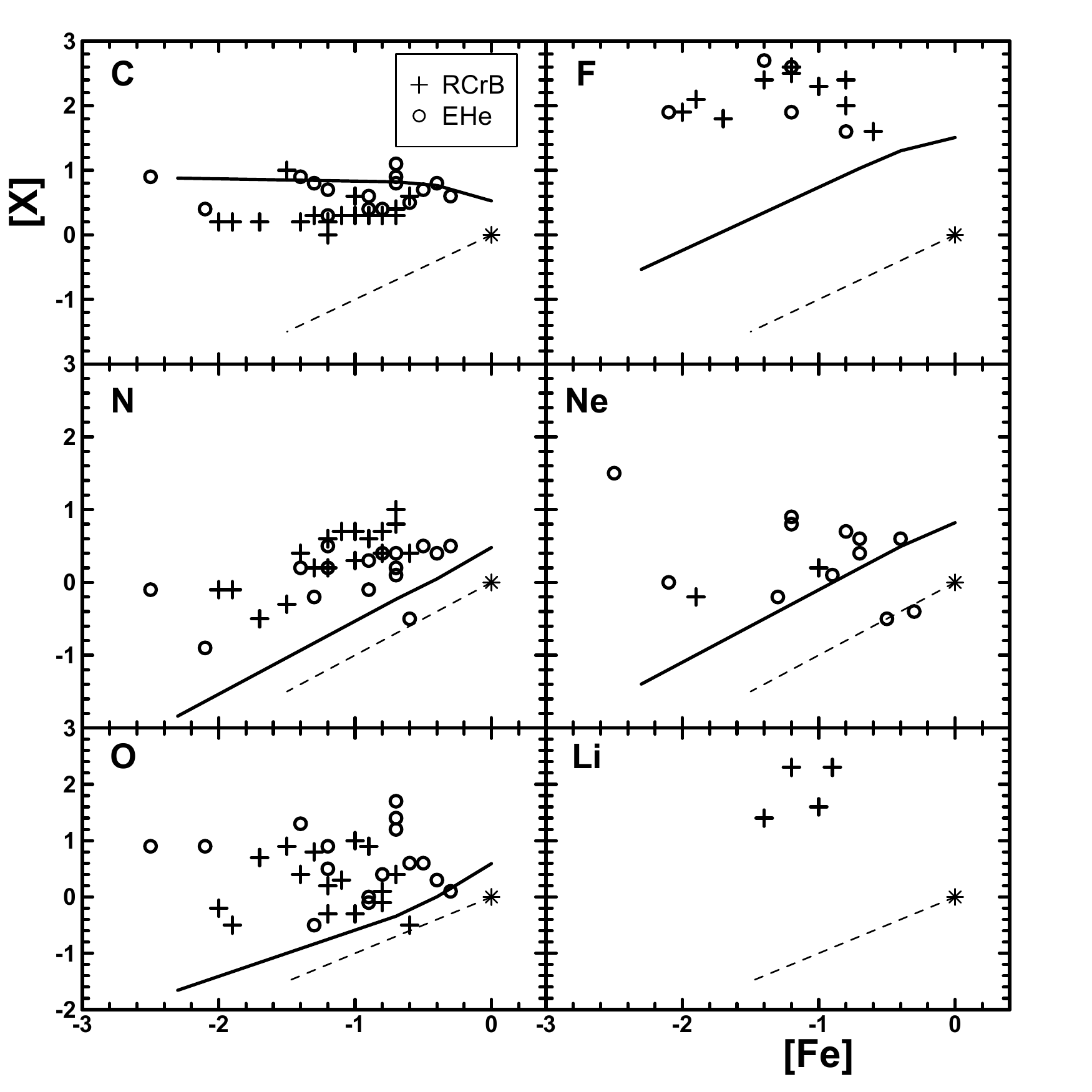} \caption{The
observed surface abundances of R\,CrB and EHe stars \citep{Jeffery11} compared with our
nucleosynthesis computation. The axes [X] and [Fe] give logarithmic
abundances relative to solar for individual elements and for iron, respectively. Solid
lines are abundances given by our simulation, which include four metallicities $z=0.0001, 0.004,
0.008$ and 0.02. The asterisks correspond to the solar composition, while the dashed diagonal
lines intersecting the solar value are the abundances expected if the solar values were
scaled with metallicity.} \label{abu} \end{figure}

The most recent measurements of the surface abundances of EHe and R CrB stars
have been collated by \citet{Jeffery11}. Here we summarise the main
observational features and compare these with our results; see Fig.~\ref{abu} for details.

\paragraph*{Carbon}  is enriched in all EHe and R\,CrB stars. Excluding MV\,Sgr,
the EHes shows a mean carbon abundance $\rm log\epsilon$=9.3, and a range from 8.9--9.7.
The carbon abundance is more difficult to measure in R\,CrBs. Furthermore, in R\,CrB stars
cool enough to show CO, the \iso{12}{C}/\iso{13}{C} ratios are very large (500),
indicating a $3\alpha$ or helium-burning origin for the carbon excess. In our models,
the \iso{12}{C} was produced by helium burning through $3\alpha$ reaction during the hot
accretion phase, and then brought up to the surface during the slow-accretion phase
by flash-driven convection. The \iso{13}{C} abundance remained almost unchanged during the whole
simulation. This enrichment of \iso{12}{C} results in a very high ratio of
\iso{12}{C}/\iso{13}{C}. The total carbon, including \iso{12}{C} and \iso{13}{C},
shows an abundance similar to the observation in Fig.~\ref{abu}.

\begin{figure} \centering \includegraphics [angle=0,scale=0.45]{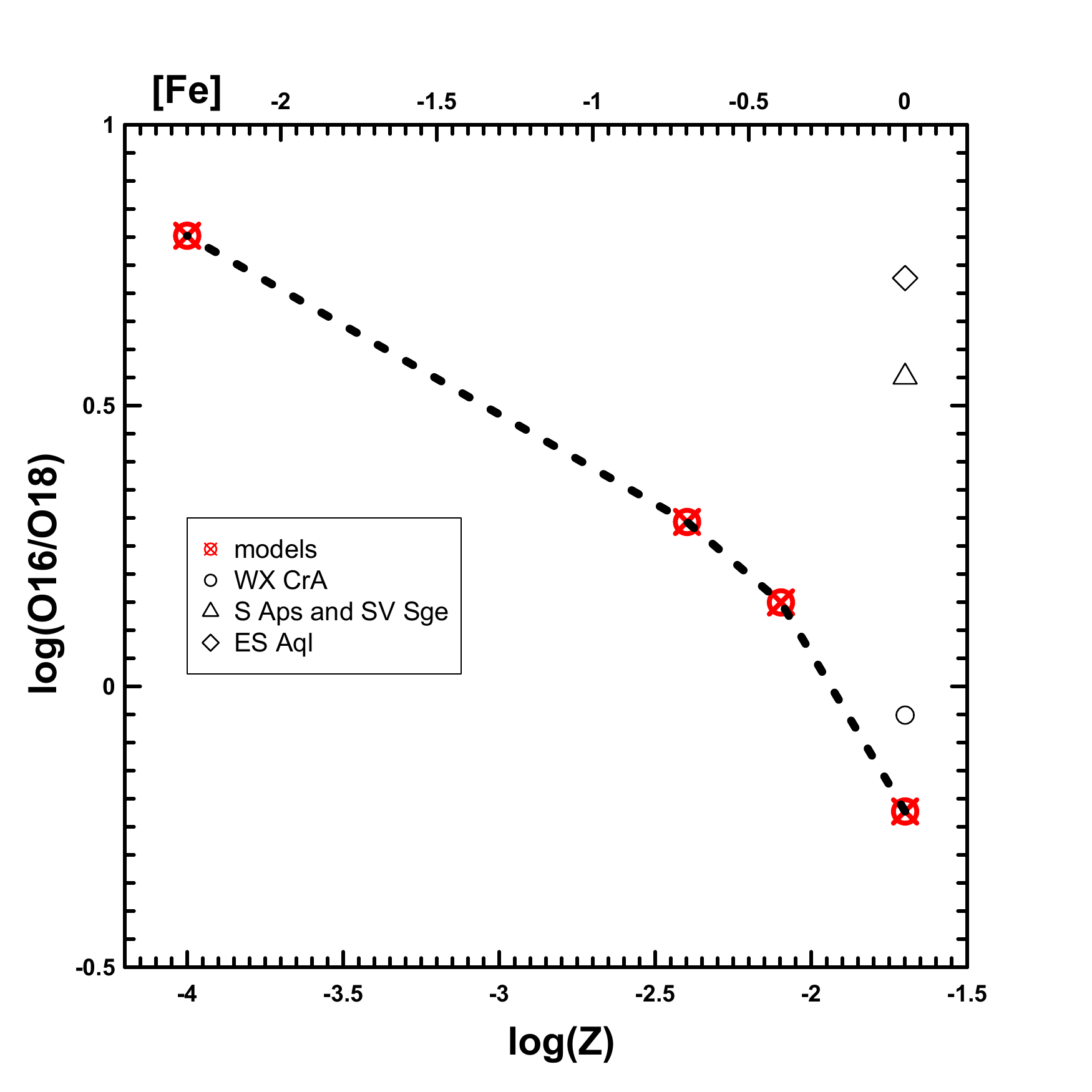}
\caption{The ratio of $\iso{16}{O}/\iso{18}{O}$ as a function of metallicity.
The computation results are shown as circles with crosses.
Diamond, triangle and circle symbols show the observations of four
$\iso{18}{O}$ enriched R\,CrB stars, i.e. WX CrA, S Aps, SV Sge and ES Aql
\citep{Clayton07}.} \label{iso} \end{figure}

\paragraph*{Nitrogen}  is enriched in the great majority of EHe and R\,CrB
stars. Heber(1983) and subsequent authors point out that the N abundances in general
follow the trend of the iron abundance. Nitrogen is enriched through CNO cycling in the parent stars.
In our simulation, it is subsequently reduced by the $\iso{14}{N}(\alpha,\gamma)\iso{18}{O} $
reaction

\paragraph*{Oxygen} is enriched in most EHe and R CrB stars.
\citet{Clayton07} report that, for four R\,CrBs, \iso{18}{O} has a large enrichment and shows a low
ratio of $\iso{16}{O}/\iso{18}{O}$ close to unity. By comparison with the solar abundance,
\iso{18}{O} must have increases by a factor $> 400$. Clayton et al.  also indicate that the production of
\iso{18}{O} requires temperatures of at least $10^8$ K to allow the
$\iso{14}{N}(\alpha,\gamma)\iso{18}{O} $ reaction. \iso{18}{O} is also produced by
$\alpha$-capture on \iso{14}{N} during the fast-accretion phase in our models;
the resulting $\iso{16}{O}/\iso{18}{O}$ ratio  is as a function of metallicity, decreasing with $Z$ (Fig.~\ref{iso}).
There is no new $\iso{16}{O}$ from alpha-capture during the fast-accretion phase.

\paragraph*{Fluorine} is enriched in several EHes \citep{Pandey06} and
in most R\,CrB stars enriched by factors of 800--8000 relative to its likely initial
abundance \citep{Pandey08}. In our simulation, \iso{19}{F}  comes from
reaction of $\iso{14}{N}(\alpha,\gamma)\iso{18}{F}(p,\alpha)\iso{15}{O}(\alpha,\gamma)\iso{19}{Ne}(\beta^+)\iso{19}{F}$.
Our simulation shows a strong overabundance of F, but not enough to fully agree with the observational data.

\paragraph*{Neon.} A high overabundance of neon has been identified in several EHes and R\,CrBs. In
our simulation, \iso{22}{Ne} is enriched by two $\alpha$-captures on \iso{14}{N} followed
by extensive convective mixing.

\paragraph*{Lithium.} A few R\,CrB stars have a notably large overabundance of lithium,
which has so far been difficult to explain. There is no significant lithium produced in our
simulation. \citet{Longland12} indicate that lithium can be produced by the
merging of a helium white dwarf with a carbon-oxygen white dwarf if their chemical
composition is rich in \iso{3}{He} from the previous evolution. This
model requires enough \iso{3}{He} to be left in the white dwarf after the end of
main-sequence evolution and a hot enough corona to form during the merger.
As a test calculation, we put
$10^{-5} $ mass fraction of  \iso{3}{He}
into the accreted material, and obtained a surface with
a lithium mass fraction of $1.5\times10^{-5}$, or about $10^5 \times$ solar.
Hence it is clear that if sufficient \iso{3}{He} is present in the accreted helium,
He+He white dwarf mergers can also yield lithium. Precisely how much depends on many factors, not least
the quantity of \iso{3}{He} in the donor and the temperature history of the fast-accretion phase.

\subsection{Some Statistics}

The merger frequency of double helium WD systems in the Galaxy is estimated to lie in the range
$0.0057 {\rm \,yr^{-1}}$ \citep{Han98} (Model 4) to $0.029 {\rm \,yr^{-1}}$ \citep{Webbink84}. \citet{Han98} (Fig.\,6)
shows the  final masses of double helium WD mergers to lie roughly in a normal distribution with a mean of
$0.61\pm 0.09 \Msolar$.  Approximately $\approx2.3\%$ have a merged total
mass $> 0.8 \Msolar$. Thus, the rate for these high-mass mergers should lie in the range
$1.3 - 6.67 \times 10^{-4} {\rm \,yr^{-1}}$.

Known low-luminosity helium stars
have effective temperatures in the range $ 4.3 \leq \log T_{\rm eff} \leq 4.5$ \citep{Jeffery96}.
Considering the evolution tracks through this temperature  range, we list the luminosity range
and evolution timescales for three different stages of evolution in Table~\ref{table:nonlinqq} (see also Fig.~\ref{gt}).
Combining the evolution time and estimates for the merger
frequency, we estimate the number of low-luminosity helium stars in the Galaxy to be
$\approx8-41$ in the inwards-burning phase (stage A), $\approx11-57$ in
the outwards-burning phase (stage B) and $\approx3-13$ cooling to the white
dwarf phase (stage C).
Taking the luminosities into account, the {\it relative} numbers of  massive double-WD mergers
observable as EHe stars in each stage  (A:B:C) should be $\approx 1:6:4$,  strongly favouring the
post-core-burning stages.

R\,CrB stars generally have effective temperatures less than 10\,000K.
In the immediate post-merger phase, our He+He merger models with $T_{\rm eff} < 10\,000K$
have a lifetime of $\approx 6.7\times10^4 \rm yrs$ and a luminosity $\approx 3$ dex
above solar. On the  post-subdwarf giant branch they have an equivalent lifetime   $\approx 1.4 \times10^4 \rm yrs$
and a luminosity $\approx$ 4 dex above solar. Thus, we can estimate the Galactic population to be
$\approx9-43$ new-born hydrogen-deficient giants from massive He+He WD mergers, and
$\approx2-9$ more luminous hydrogen-deficient giants from post-core-burning He-sdO
stars. Assuming no intrinsic extinction, the relative numbers of observable cool H-deficient giants in
each phase from this channel should be $\approx 1:3$.

Finally, if we assume a rate of $0.018-0.019 {\rm \,yr^{-1}}$ for the merger of CO+He white dwarfs \citep{Webbink84,Han98},
and a mean evolution timescale which is shorter by a factor of 2 due to their higher masses,
there would be  $\approx14-70$ times as many R\,CrB stars from the CO+He merger channel as from the massive  He+He channel.

    \begin{table}

\caption{Post-merger models with $4.3 \leq \log T_{\rm eff} \leq 4.5$. }
 \centering \begin{tabular}{l l l l }
  \hline
  \hline
  Stage &$\log L/{\rm L_{\odot}}$ &Timescale (yrs) & Numbers\\ [0.5ex]	
   \hline
    A & $2.7-3$ &$ 6\times10^4 $  & $8 - 41$   \\
    B & $3.6-3.7$& $8.6\times10^4 $  & $11 - 57$  \\
    C & 4 & $ 2\times10^4 $  & $3 - 13$  \\
    \hline
    \end{tabular}
    \label{table:nonlinqq}
    \end{table}

\section{Conclusion}

Mergers of helium and carbon-oxygen white dwarfs are currently regarded to be the
most favoured model for the origin of most R\,CrB and EHe stars. The merger of
two helium white dwarfs has been widely considered responsible for the
origin of some hot subdwarfs, particulalry those with hydrogen-deficient surfaces.
In this paper we have demonstrated that the most massive helium+helium white dwarf mergers
can produce hot subdwarfs which subsequently become cool supergiants and, consequently,
provide an alternative means to produce low-mass R\,CrB and EHe stars. However, the number of R\,CrB stars
 produced in this way may be  some 14 to 70 times smaller than from the CO+He WD merger channel.
Nucleosynthesis of elements during the hot (fast-accretion) phase of the merger is roughly consistent
with the observed abundances of  \iso{12}{C}, \iso{18}{O}, \iso{19}{F} and \iso{22}{Ne} in  R\,CrB and EHe stars.
Further work is required to establish whether the products of massive He+He WD mergers can be clearly distinguished
from the products of CO+WD mergers.

\section*{Acknowledgments} The Armagh Observatory is supported by a grant from the
Northern Ireland Dept. of Culture Arts and Leisure.

\bibliographystyle{mn} 
\bibliography{mybib} 

\end{document}